\documentclass{article}
\usepackage{ismir,amsmath,cite,url}
\usepackage{graphicx}
\usepackage{color}
\usepackage{multirow}
\usepackage[symbol]{footmisc}

\title{END-TO-END LEARNING FOR MUSIC AUDIO TAGGING AT SCALE}

\oneauthor
 {\hspace{-2mm}Jordi Pons$^\star$$^\dagger$ \hspace{0.05mm} Oriol Nieto$^\dagger$ \hspace{0.05mm} Matthew Prockup$^\dagger$ \hspace{0.05mm} Erik Schmidt$^\dagger$ \hspace{0.025mm} Andreas Ehmann$^\dagger$ \hspace{0.025mm} Xavier Serra$^\star$}
 {$^\star$ Music Technology Group, Universitat Pompeu Fabra, Barcelona\\
 $^\dagger$ Pandora Media Inc., Oakland, CA\vspace{-3mm}}

\begin{document}

\maketitle
\begin{abstract}
The lack of data tends to limit the outcomes of deep learning research, particularly when dealing with end-to-end learning stacks processing raw data such as waveforms.\linebreak In this study, 1.2M tracks annotated with musical labels are available to train our end-to-end models. This large amount of data allows us to unrestrictedly explore two different design paradigms for music auto-tagging: assumption-free models -- using waveforms as input with very small convolutional filters; and models that rely on domain knowledge~-- log-mel spectrograms with a convolutional neural network designed to learn timbral and temporal features. Our work focuses on studying how these two types of deep architectures perform when datasets of variable size are available for training: the MagnaTagATune (25k songs), the Million Song Dataset (240k songs), and a private dataset of 1.2M songs.
Our experiments suggest that music domain assumptions are relevant when not enough training data are available, thus showing how waveform-based models outperform spectrogram-based ones in large-scale data scenarios.

\end{abstract}
\vspace{-3mm}
\section{Introduction}

One fundamental goal in music informatics research is to automatically structure large music collections.
The music audio tagging task consists of automatically estimating the musical attributes of a song --
including: moods, language of the lyrics, year of composition, genres, instruments, harmony, or rhythmic traits.
Thus, tag estimates may be useful to define a semantic space that can be advantageous for automatically organizing musical libraries.

Many approaches have been considered for this task (mostly based on feature extraction + model~\cite{bayle2017revisiting,sordo2007annotating,matt}), with recent publications showing promising results using deep architectures 
\cite{dieleman2014end,choi2016automatic,pons2017timbre,lee2017sample}. In this work we confirm this trend by studying how two deep architectures conceived considering opposite design strategies (using domain knowledge or not) perform for several datasets -- with one of the datasets being of an unprecedented size: 1.2M songs. 
Provided that a sizable amount of data is available for that study, we investigate the learning capabilities of these two architectures. 
Specifically, we investigate whether the architectures based on domain knowledge overly constrain the solution space for cases where large training data are available -- in essence, we study if certain architectural choices (e.g., using log-mel spectrograms as input) can limit the model's capabilities to learn from data.
The main contribution of this work is to show that little to no model assumptions are required for music auto-tagging when operating with large amounts of data.

Section 2 discusses the main deep architectures we identified in the audio literature, section~3 describes the datasets used for this work, section~4 presents the architectures we study, and section~5 provides discussion about the results with conclusions drawn in section 6.
\vspace{-3mm}
\section{Current deep architectures}

\begin{figure}[h]
	\centering
	\includegraphics[width=1\linewidth]{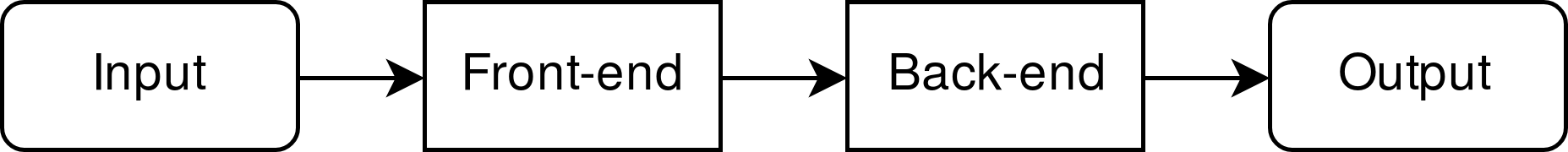}
	\vspace{-7mm}
	\caption{Deep learning pipeline.}
	\label{fig:parts}
\end{figure}

In order to facilitate the discussion around the current audio architectures, we divide deep learning models into two parts: front-end and back-end -- see Figure \ref{fig:parts}. The front-end is the part of the model that interacts with the input signal in order to map it into a latent-space, and the back-end predicts the output given the representation obtained by the front-end. In the following, we present the main front- and back-ends we identified in the literature.
\vspace{-1mm}

\vspace{-3mm}

\textbf{Front-ends.}
These are generally comprised of convolutional neural networks (CNNs) \cite{dieleman2014end,choi2016automatic,zhu2016learning,pons2017designing,pons2017timbre}, since these can learn efficient representations by sharing weights\footnote{Which determine the learned feature representations.}along the signal. Front-ends can be divided into two groups depending on the used input signal: waveforms \cite{dieleman2014end,zhu2016learning,lee2017sample} or spectrograms \cite{choi2016automatic,pons2017designing,pons2017timbre}.
Further, the design of the filters can be either based on domain knowledge or not. For example, one leverages domain knowledge when a front-end for waveforms is designed so that 
the length of the filter is set to be as the window length of a STFT \cite{dieleman2014end}. Or for a spectrogram front-end, it is used vertical filters to learn timbral representations~\cite{lee2009unsupervised} or~horizontal filters to learn longer temporal cues~\cite{schluter2014improved}. 
Generally, a single filter shape is used in the first CNN layer \cite{dieleman2014end,choi2016automatic,lee2009unsupervised,schluter2014improved}, but some recent works report performance gains when using several filter shapes in the first layer \cite{zhu2016learning,pons2017designing,pons2017timbre,phan2016robust,pons2016experimenting,chenhigh}. Using many filters promotes a richer feature extraction in the first layer, and facilitates leveraging domain knowledge for designing the filters' shape. 
For example: a waveform front-end using many long filters (of different lengths) can be motivated from the perspective of a multi-resolution time-frequency transform\footnote{The Constant-Q Transform \cite{brown1991calculation} is an example of such transform.}\cite{zhu2016learning}; or since it is known that some patterns in spectrograms are occurring at different time-frequency scales, one can intuitively incorporate many (different) vertical and/or horizontal filters in a spectrogram front-end \cite{pons2017designing,pons2017timbre,pons2016experimenting,phan2016robust}.
To summarize, using domain knowledge when designing models allows us to naturally connect the deep learning literature with previous signal processing work.
On the other hand, when domain knowledge is not used, it is common to employ a deep stack of small filters, e.g.: 3$\times$1 as in the sample-level front-end used for waveforms~\cite{lee2017sample}, or 3$\times$3 filters used for spectrograms~\cite{choi2016automatic}. These models based on small filters make minimal assumptions over the local stationarities of the signal, so that any structure can be learned via hierarchically combining small-context representations. 
These architectures with small filters are flexible models able to potentially learn any structure given enough depth and data.

\textbf{Back-ends.}
Among the different back-ends used in the audio literature, we identified two main groups: \textit{(i)}~fixed-length input back-end, and \textit{(ii)}~variable-length input back-end. The generally convolutional nature of the front-end allows it to process different input lengths. Therefore, the back-end unit can adapt a variable-length feature map 
to a fix-sized output. The former group of models (\textit{i}) assume that the input will be kept constant -- examples of those are front-ends based on feed-forward neural-networks or fully-convolutional stacks~\cite{dieleman2014end,choi2016automatic}. The second group (\textit{ii}) can deal with different input-lengths since the model is flexible in at least one of its input dimensions -- examples of those are back-ends using temporal-aggregation strategies such as max-pooling, average-pooling, attention models or recurrent neural networks \cite{collin}. Given that songs are generally of different lengths, 
these types of back-ends are ideal candidates for music processing. However, despite the different-length nature of music, many works employ fixed-length input back-ends (group \textit{i}) since these architectures tend to be simpler and perform well \cite{choi2016automatic,pons2017timbre,dieleman2014end}.

\begin{figure*}[t]
	\centering
	\includegraphics[width=1\linewidth]{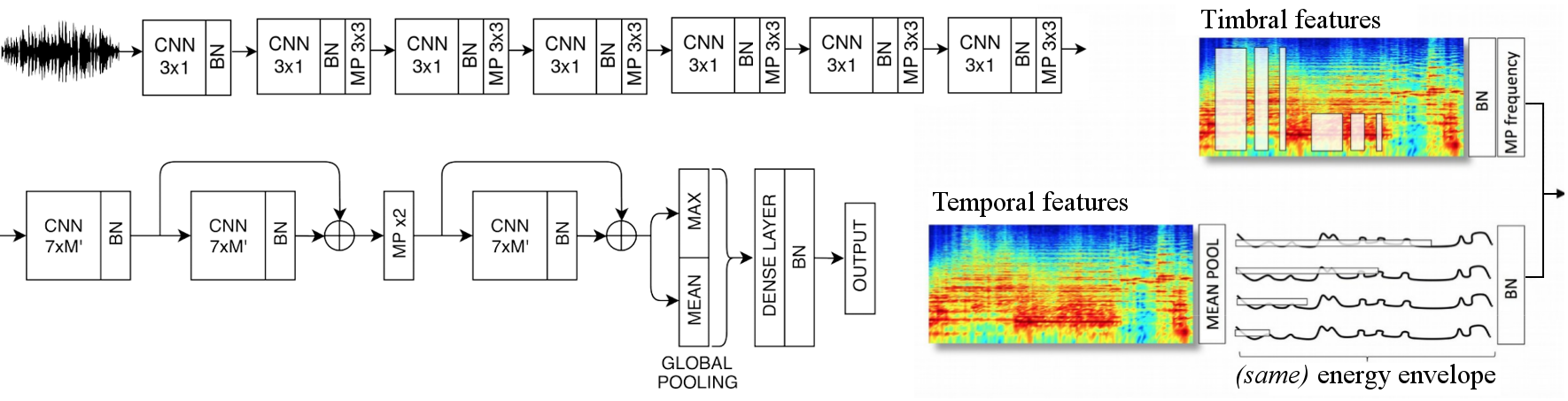}
		\vspace{-6mm}
	\caption{\textit{\textbf{Bottom-left}} -- back-end. \textit{\textbf{Top-left}} -- waveform front-end. \textit{\textbf{Right}} -- spectrogram front-end. \textbf{\textit{Definitions}} -- \textit{M'} stands for the feature map's vertical axis, \textit{BN} for batch norm, and \textit{MP} for max-pool. }
	\vspace{-4.1mm}
	\label{fig:backend}
\end{figure*}

\vspace{-4mm}
\section{DATASETS}
 \vspace{-1mm}


We study how different deep architectures for music auto-tagging perform for 3 music collections of different sizes:



1) The \underline{MagnaTagATune} (MTT) dataset is of $\approx$~26k music audio clips of 30s\cite{law2009evaluation}.
Predicting the top-50 tags of this dataset is a popular benchmark for auto-tagging. 

2) Although the \underline{Million Song Dataset} (MSD) name indicates that 1M songs are available\cite{bertin2011million}, audio files with proper tag annotations (top-50 tags) are only available for~$\approx$ 240k previews of 30s. This dataset constitutes the biggest public dataset available for music auto-tagging, making these data highly appropriate for benchmarking.

3) A private dataset consisting of 1M songs for training, 100k for validation, and 100k for test\footnote{Test \& validation sets are kept the same throughout the experiments for a fair evaluation. All used partitions are stratified and artist-filtered.}is available for this study. The \underline{1.2M-songs} dataset has 139 track-level human-expert annotations that can be summarized as follows:

$\cdot$ \textit{Meter tags} denote different sorts of musical meters ({e.g.}, triple-meter, cut-time, compound-duple, odd).

$\cdot$ \textit{Rhythmic feel tags} denote rhythmic interpretation ({e.g.},~swing, shuffle, back-beat strength) and elements of rhythmic perception (\textit{e.g}., syncopation, danceability).

$\cdot$ \textit{Harmonic tags}: major, minor, chromatic, etc.

$\cdot$ \textit{Mood tags} express the sentiment of a music audio clip ({e.g.}, if the music is angry, sad, joyful).

$\cdot$ \textit{Vocal tags} denote the presence of vocals and timbral characteristics of it ({e.g.}, male, female, vocal grittiness).

$\cdot$ \textit{Instrumentation tags} denote the presence of instruments ({e.g.}, piano) and their timbre ({e.g.}, guitar distortion).

$\cdot$ \textit{Sonority tags} detail production techniques ({e.g.},~studio, live) and overall sound (e.g., acoustic, synthesized).

$\cdot$ \textit{Basic genre tags}: jazz, rock, rap, latin, disco, etc.

$\cdot$ \textit{Subgenre tags}: jazz (e.g., cool, fusion, hard bop), rock (e.g., light, hard, punk), rap (e.g., east coast, old school), world music (e.g., cajun, indian), classical music (e.g.,~baroque period, classical period), etc.

\vspace{1mm}
\noindent Other large (music) audio datasets exist: the Free Music Archive (FMA: $\approx$106k songs)\cite{defferrard2017fma} and Audioset ($\approx$2.1M audios)\cite{gemmeke2017audio}. 
Since previous works mainly used the MTT and MSD\cite{choi2016automatic,lee2017sample}, we employ these datasets to assess the studied models with public data. 
Despite our interest in using FMA, for brevity, we restrict our study to 3 datasets that already cover a wide range of different sizes. Finally, Audioset is not used since most of its content is not music.

\vspace{-4mm}
 
 \section{THE ARCHITECTURES UNDER STUDY}
 \vspace{-1mm}
 
After an initial exploration of the different architectures introduced in section 2, we select two models based on opposite design paradigms: one for processing waveforms, with a design that does minimal assumptions over the task at hand; and another for spectrograms, with a design that heavily relies on musical domain knowledge. 
Our goal is to compare these two models for providing insights in whether domain knowledge is required (or not) for designing deep learning models.
This section provides discussion around our architectural choices and introduces the \textit{basic} configuration setup -- which is also accessible online.\footnote{https://github.com/jordipons/music-audio-tagging-at-scale-models}

The waveform model was selected after observing that the sample-level front-end (using a deep stack of 3$\times$1 filters) was remarkably superior to the other waveform-based front ends -- as shown in the original paper~\cite{lee2017sample}. This result is particularly compelling because this front-end does not rely on domain-knowledge for its design. Note that raw waveforms are fed to the model without any pre-processing, and the small filters considered for its design make no strong assumptions over the most informative local stationarities in waveforms. Therefore, the sample-level can be seen as a problem agnostic front-end that has the potential to learn any audio task provided that enough depth and data are available. Given that a large amount data is available for this study, the sample-level front-end is of particular interest due to its strong learning potential: its solution space is not constrained by severe architectural choices relying on domain knowledge.


On the other hand, when experimenting with spectrogram front-ends, we found domain knowledge intuitions to be valid guides for designing deep architectures.
For example, front-ends based on \textit{(i)} many vertical and horizontal filters in the first layer were consistently superior to front-ends based on \textit{(ii)} a single vertical filter~-- as shown in recent publications \cite{chenhigh,pons2017designing,pons2016experimenting,phan2016robust}.
Note that the former front-ends \textit{(i)} can learn spectral and (long) temporal representations already in the first layer -- which are known to be important musical cues; while the latter \textit{(ii)} can only learn spectral representations.
Moreover, we observed that front-ends based on a deep stack of 3$\times$3 filters were achieving equivalent performances to the former front-end \textit{(i)} when input segments were shorter than 10s -- as noted in the literature\cite{pons2017timbre}. But when considering longer inputs (which yielded better performance), the computational price of this deeper model increases: longer inputs implies having larger feature maps in every layer and therefore, more GPU memory consumption. For that reason, we refrained from using a deep stack of 3$\times$3 filters as a front-end -- because our 12GBs of VRAM were not enough to input 15s of audio when using a back-end. 
Hence, making use of domain knowledge  also provides guidance for minimizing the computational cost of the model -- since by using a single layer with many vertical and horizontal filters, one can efficiently capture the same receptive field without paying the cost of going deep. 
Finally, note that front-ends using many vertical and horizontal filters in the first layer are an example of deep architectures relying on (musical) domain knowledge for their design.


 

After considering the previous discussion, we select the sample-level front-end as main part of our assumption-free model for waveforms; and we use a spectrogram front-end with many vertical and horizontal (first-layer) filters for the model designed considering domain knowledge.
Experiments below share the same back-end, which enables a fair comparison among the previously selected front-ends.
Unless otherwise stated, the following specifications are the ones used for the experiments -- throughout the document, we refer to these specifications as the \textit{basic} configuration:

\textbf{Shared back-end.}
It consists of three CNN layers (with 512 filters each and two residual connections), two pooling layers and a dense layer -- see Figure \ref{fig:backend} (\textit{Bottom-left}). 
We introduced residual connections in our model to explore very deep architectures, such that we can take advantage of the large data available. 
Although adding more residual layers did not drastically improve our results, we observed that adding these residual connections stabilized learning while slightly improving performance~\cite{li2017visualizing}. 
The used 1D-CNN filters~\cite{dieleman2014end} are computationally efficient and shaped such that all extracted features are considered across a reasonable amount of temporal context (note the 7$\times$M' filter shapes, representing \textit{time}$\times$\textit{all~features}). 
We also make a drastic use of temporal pooling: firstly, down-sampling \textit{x2} the temporal dimensionality of the feature maps; and secondly, by making use of global pooling with mean and max statistics. The global pooling strategy allows for variable length inputs to the network and 
therefore, such a model can be classified as a ``variable-length input" back-end.
Finally, a dense layer with 500 units connects the pooled features to a sigmoidal output. 

\textbf{Waveform front-end.}
It is based on a sample-level front-end \cite{lee2017sample} composed of seven: 1D-CNN (3$\times$1 filters), batch norm, and max pool layers -- see Figure \ref{fig:backend} (\textit{Top-left}).\linebreak Each layer has 64, 64, 64, 128, 128, 128 and 256 filters.
For the 1.2M-songs dataset, we use a model with more capacity having nine layers with 64, 64, 64, 128, 128, 128, 128, 128, 256 filters.
By hierarchically combining small-context representations and making use of max pooling, the sample-level front-end yields a feature map for an audio segment of 15s (down-sampled to 16kHz) which is further processed by the previously described back-end.

\textbf{Spectrogram front-end.} 
Firstly, audio segments are converted to log-mel magnitude spectrograms (15 seconds and 96 mel bins \cite{peeters2004large}) and normalized to have zero-mean and unit-var. Secondly, we use vertical and horizontal filters explicitly designed to facilitate learning the timbral and temporal patterns present in spectrograms \cite{pons2016experimenting,pons2017designing,pons2017timbre}.
Note in Figure~\ref{fig:backend}~(\textit{Right}) that the spectrogram front-end is a single-layer CNN with many filter shapes that are grouped into two branches\cite{pons2016experimenting}: \textit{(i}) top branch~-- timbral features \cite{pons2017timbre}; and \textit{(ii)} lower branch -- temporal features~\cite{pons2017designing}. 
The top branch is designed to capture pitch-invariant timbral features that are occurring at different time-frequency scales in the spectrogram. Pitch invariance is enforced via enabling CNN filters to convolve through the frequency domain, and via max-pooling the feature map across its vertical axis \cite{pons2017timbre}. Note that several filter shapes are used to efficiently capture many different time-frequency patterns: 7$\times$86, 3$\times$86, 1$\times$86, 7$\times$38, 3$\times$38 and 1$\times$38\footnote{Each filter shape has 16, 32, 64, 16, 32 and 64 filters, respectively.}-- to facilitate learning, e.g.: kick-drums (with small-rectangular filters of 7$\times$38 capturing sub-band information for a short period of time), or string ensemble instruments (with long vertical filters of 1$\times$86 which are capturing timbral patterns spread in the frequency axis). The lower branch is meant to learn temporal features, and is designed to efficiently capture different time-scale representations by using several long filter shapes~\cite{pons2017designing}: 165$\times$1, 128$\times$1, 64$\times$1 and 32$\times$1.\footnote{Each filter shape has 16, 32, 64 and 128 filters, respectively.}These filters operate over an energy envelope (not directly over the spectrogram) obtained via mean-pooling the frequency-axis of the spectrogram. 
By computing the energy envelope in that way, we are considering high and low frequencies together while minimizing the computations of the model~-- note that no frequency/vertical convolutions are performed, but 1D (temporal) convolutions. Thus, 
domain knowledge is also providing guidance to minimize the computational cost of the model. The output of these two branches is merged, and the previously described back-end is used for going deeper. For further details, see its online implementation.$^4$

\textbf{Parameters.}
50\% dropout before every dense layer, \mbox{ReLUs} as non-linearities, and our models are trained with SGD employing Adam (with an initial learning rate of~0.001) as optimizer. We minimize the MSE for the 1.2M-songs dataset, but we minimize the cross entropy for the other datasets. During training our data are converted to audio patches of 15s, but during prediction one aims to consider the whole song. To this end, several predictions are computed for a song (by a moving window of 15s) and then averaged. Although our models are capable of predicting tags for variable-length inputs, we use fixed length patches since in preliminary experiments we observed that predicting the whole song at once yielded worse results than averaging several patch predictions. In future work we aim to further study this behavior, to find ways to exploit the fact that the whole song is generally available.

\begin{table}[]
	\begin{center}
		\begin{tabular}{l | c | c  c  c }
			\textbf{1.2M-songs}&\textit{train} & \textit{ROC} & \textit{PR} & \\			
			\textit{Models} &\textit{size} & \textit{AUC} & \textit{AUC} & $\mathit{\sqrt{MSE}}$  \\
			\hline \hline
			Baseline &1.2M & 91.61\%& 54.27\%& 0.1569 \\ \hline
			Waveform &1M & \textbf{92.50}\%& \textbf{61.20}\%& \textbf{0.1465} \\
			Spectrogram & 1M & {92.17\%}& {59.92\%}& {0.1473} \\ \hline
			Waveform & 500k & 91.16\%& 56.42\%& 0.1504 \\
			Spectrogram & 500k & 91.61\%& 58.18\%& 0.1493\\ \hline
			Waveform & 100k & 90.27\%& 52.76\%& 0.1554 \\
			Spectrogram &100k & 90.14\%& 52.67\%& 0.1542\\ 
		\end{tabular}
	\end{center}

	\vspace{-4mm}
		\caption{1.2M-songs average results (3 runs) when using different training-set sizes. Baseline: GBTs+features \cite{matt}.}
	\label{tab:example}
	\vspace{-4mm}
\end{table}
\vspace{-3mm}
\section{Experimental results}
\vspace{-1mm}

\vspace{-1mm}
\subsection{1.2M-songs dataset}
\vspace{-1mm}
\hspace{4mm}\textbf{Experimental setup.} As a baseline, we use a system consisting of a music feature extractor (in essence: timbre, rhythm, and harmony descriptors) and a model based on gradient boosted trees (GBT) for predicting each of the tags \cite{matt}. By predicting each tag individually, one aims to turn a hard problem into multiple (hopefully \textit{simpler}) problems. 
A careful inspection of the dataset reveals that, among tags, two different data distributions dominate the annotations: \textit{(i)} tags with bi-modal distributions, where most of the annotations are zero, which can be classified; and \textit{(ii)}~tags with pseudo-uniform distributions that can be regressed.\footnote{Note that all output nodes are sigmoidal -- i.e., we treat classification tags as regression tags for simplicity's sake.}A regression tag example is \textit{acoustic}, which indicates how acoustic a song is -- from zero to one, zero being an electronic music song and one a string quartet. And a classification tag example can be any genre -- for example, most songs will not be cataloged as \textit{rap} since the dataset is large and its taxonomy contains dozens of genres.
We use two sets of performance measurements: ROC-AUC\footnote{ROC: Receiver Operating Characteristic. PR: Precision Recall. AUC: Area Under the Curve. MSE: Mean Squared Error.}and PR-AUC\footnotemark[8]for the classification tags, and error ($\sqrt{MSE}$\footnotemark[8]) for the regression tags. ROC-AUC can lead to over-optimistic scores in cases where data are unbalanced \cite{davis2006relationship}; given that classification tags are highly unbalanced, we also consider the PR-AUC metric since it is more indicative than ROC-AUC in these cases \cite{davis2006relationship}.
For ROC-AUC and PR-AUC, the higher the score the better -- but for $\sqrt{MSE}$, the lower the better.
Studied spectrogram and waveform models are set following the \textit{basic} configuration -- and are composed of 5.9M and 5.5M parameters, respectively.
Given the unprecedented size of the dataset, we focus on how these models scale when trained with different amounts of data: 100k, 500k, or 1M songs.
Average results (across 3 runs) are shown in Table~\ref{tab:example} and Figure 3.

\begin{figure}[t]
	\centering
	\includegraphics[width=0.87\linewidth]{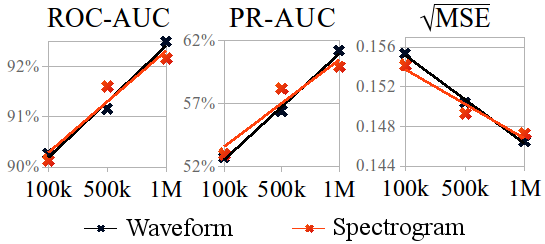}
		\vspace{-4mm}
	\caption{Linear regression fit on the 1.2M-songs results.}
	\label{fig:graphic}
		\vspace{-5mm}
\end{figure}


\textbf{Quantitative results.} 
Training the models with 100k songs took a few days, with 500k songs one week, and with 1M songs less than two weeks.
The deep learning models trained with 1M tracks achieve better results than the baseline in every metric. However, the deep learning models trained with 100k tracks perform worse than the baseline.
This result confirms that deep learning models require large datasets to clearly outperform strong 
methods based on feature-design -- although note that large datasets are generally not available for most audio tasks. Moreover, the biggest performance improvement \textit{w.r.t.} the baseline is seen for PR-AUC, which provides a more informative picture of the performance when the dataset is unbalanced~\cite{davis2006relationship}. 
In addition, the best performing model is based on the waveform front-end -- being capable of outperforming the spectrogram model in every metric when trained with 1M songs. 
This result confirms that waveform sample-level front-ends have a great potential to learn from large data, since their solution space is not constrained by any severe architectural choice. On the other hand, the architectural choices defining the spectrogram front-end might be constraining the solution space. While these architectural constraints are not harmful when training data are scarce \linebreak(as~for the 100k/500k songs results or in prior works \cite{sainath2015learning}), such a strong regularization of the solution space may limit the learning capacity of the model in scenarios where large training data are available -- as~for the 1M songs results.
One can observe this in Figure 3, where we fit linear models to the obtained results to further study this behavior. When 100k training songs are available: trend lines show that spectrogram models tend to perform better. However, when 1M training songs are available: the lines show that waveform models outperform the spectrogram ones. It is worth mentioning that the observed trends are consistent throughout metrics: ROC-AUC, PR-AUC, and $\sqrt{MSE}$.
Finally, note that there is room for improving the models under study -- e.g.: one could address the data imbalance problem during training, or improve the back-end via exploring alternative temporal aggregation strategies.

\textbf{Qualitative results.} Since it is the first report of a deep music tagging model trained with such a large dataset, we also perceptually assess the quality of the estimates. To this end, we compared the predictions of one of our best performing models to the predictions of the baseline, and to the human-annotated ground-truth tags. Some interesting examples identified during this qualitative experiment are available online.\footnote{http://www.jordipons.me/apps/music-audio-tagging-at-scale-demo}First, we observed that the deep learning model is biased towards predicting the popular tags (such as \textit{‘lead vocals’}, \textit{‘English’} or \textit{‘male vocals’}). Note that this is expected since we are not addressing the data unbalancing issue during training. And second, we observe that the baseline model (which predicts the probability of each tag with an independent GBT model) predicts mutually exclusive tags with high confidence --
e.g., it predicted with high scores: \textit{East Coast} and \textit{West Coast} for an East Cost rap song, or \textit{baroque period} and \textit{classic period} for a Bach aria. However, the deep learning model (predicting the probability of all tags together) was able to better differentiate these similar but mutually exclusive tags. This suggests that deep learning has an advantage when compared to traditional approaches, since these mutually exclusive relations can be jointly encoded within the~model.
\vspace{-3mm}
\subsection{MagnaTagATune (MTT) dataset}
\vspace{-1mm}
\hspace{4mm} \textbf{Experimental setup.} State-of-the-art models are set as baselines, and we use the same (classification) performance metrics as for the 1.2M-songs dataset: ROC-AUC and PR-AUC -- note that the MTT labels are binary.\linebreak
One of the baseline results (the SampleCNN \cite{lee2017sample} with 90.55 ROC-AUC) was computed using a slightly different version of the MTT dataset -- which only includes songs having more than 1 tag and lasting more than 29.1 seconds. As a result, this cleaner version of the MTT dataset is of $\approx$21k songs instead of $\approx$26k. Although this dataset cleans out potential noisy annotations, we decided to use the original dataset to easily compare our results with former works. Thus, to fairly compare our models with the SampleCNN, we reproduce their work considering the original dataset -- achieving a score of 88.56 ROC-AUC. Given that less noise is present in the SampleCNN dataset, it seems reasonable that their performance is higher than the one obtained by our implementation.

The MTT experiments can be divided in two parts: waveform and spectrogram models -- see Tables \ref{tab:wave} and~\ref{tab:spec}. Due to the amenable size of the dataset (every MTT experiment lasts $<$ 5h), it is feasible to run a comprehensive study investigating different architectural configurations. Specifically, we study how waveform and spectrogram architectures behave when modifying the capacity of their front- and back-ends. For example, the experiment\linebreak``\# filters $\times$1/2" in Table \ref{tab:wave} consists of dividing the number of filters available in the waveform front-end by two. This means having 32,
32, 32, 64, 64, 64 and 128 filters, instead of the 64, 64, 64, 128, 128, 128 and 256 filters in the \textit{basic} configuration.
We also apply this methodology to the spectrogram front-ends, and we add/remove  capacity to them by increasing/decreasing the number of available filters. After running the front-end experiments with a fixed back-end (following the \textit{basic} configuration: 512 CNN filters, 500 output units), we select the most promising ones to proceed with the back-end study~-- for~waveforms: ``\#~filters~$\times$2",\footnote{``\# filters $\times$2" front-end was selected instead of ``\# filters $\times$4", because it performs similarly with less parameters.}and for spectrograms: \linebreak``\# filters~$\times$1/2". Having now a fixed front-end for every experiment, we modify the capacity of the back-end via changing the number of filters in every CNN layer (512, 256, 128, 64) and changing the number of output units (500, 200). Since the \textit{basic} configuration leads to relatively big models for the size of the dataset, these experiments explore smaller back-ends. The inputs for the MTT are set to be of 3s, since longer inputs yield worse results\cite{pons2017timbre,lee2018samplecnn}.

\textbf{Quantitative results.} 
The waveform and spectrogram models we study outperform the proposed baselines -- which represent the current state-of-the-art.
Further, performance is quite robust to the number of parameters of the model. Although the best results are achieved by models having higher capacity, the performance difference between small and large models is minor -- what means that relatively small models (which are easier to deploy) can do a reasonable job when tagging the MTT music.
Finally: spectrogram models perform better than waveform models for this small public dataset~-- which aligns with previous works using datasets of similar size~\cite{pons2017timbre,pons2017designing}. Consequently, these results confirm that domain knowledge intuitions are valid guides for designing deep architectures in scenarios where training data are scarce.

\begin{table}[]
	\centering
	\begin{tabular}{l|c|c|c}
\textbf{MTT dataset} & \textit{ROC} & \textit{PR} & \# \\ 
\textit{Waveform models} & \textit{AUC} & \textit{AUC} & \textit{param} \\
\multicolumn{4}{l}{\textit{}} \\
\multicolumn{4}{l}{\textit{State-of-the-art results -- with our own implementations}} \\ \hline
SampleCNN\cite{lee2017sample}\footnotemark[13]	& 90.55 & - & 2.4M
 \\
SampleCNN {(reproduced)} & 88.56 & 34.38 & 2.4M\\
Dieleman et al.\cite{dieleman2014end}  & 84.87& - & - \\
Dieleman et al. {(reproduced)} & 85.58 & 29.59 & 194k\\
\multicolumn{4}{l}{\textit{}} \\
\multicolumn{4}{l}{\textit{How much capacity is required for the front-end?}} \\
\hline
\# filters $\times$4 & \textbf{89.05} & \textbf{34.92} & 11.8M \\
\# filters $\times$2 \hspace{6mm}(selected) & {88.96} & {34.74} & 7M \\
\# filters $\times$1 & 88.9 & 34.18& 5.3M \\
\# filters $\times$1/2& 88.69 & 33.97& 4.7M \\
\# filters $\times$1/4& 88.47 & 33.89 & 4.4M \\
\multicolumn{4}{l}{\textit{}} \\
\multicolumn{4}{l}{\textit{How much capacity is required for the back-end?}} \\
\multicolumn{4}{l}{\# filters in every CNN layer - \# units in dense layer} \\
\hline
\hspace{1.2mm} 64 CNN filters - 500 units  & 88.57 & 33.99 & 1.3M\\
\hspace{23.5mm}- 200 units  & \underline{88.94}&\underline{34.47} & 1.3M\\
128 CNN filters - 500 units  & 88.82 & 34.62 & 1.8M \\
\hspace{22.5mm} - 200 units  & 88.81 & 34.6 & 1.7M\\
256 CNN filters - 500 units  & 88.95 & 34.27 & 3.1 M\\
\hspace{22.5mm} - 200 units  &88.59&34.39 & 2.9M\\
512 CNN filters - 500 units &  \underline{88.96}&\underline{34.74}& 7M \\
\hspace{22.5mm} - 200 units  &  88.3&34.05 & 6.7M\\

	\end{tabular} 
		\vspace{-2mm}
			\caption{MTT results: waveform models.}
		\label{tab:wave}
					\vspace{-3mm}
\end{table}

\begin{table}[t]
	\centering
	\begin{tabular}{l|c|c|c}
\textbf{MTT dataset} & \textit{ROC} & \textit{PR}& \# \\ 
\textit{Spectrogram models}  & \textit{AUC} & \textit{AUC}& \textit{param} \\
\multicolumn{4}{l}{\textit{}} \\
\multicolumn{4}{l}{\textit{State-of-the-art results -- with our own implementations}} \\ \hline
VGG - Choi et al. \cite{choi2016automatic}	& 89.40 & - & 22M\\
VGG {(reproduced)}	& 89.99 & 37.56 & 450k\\
Timbre CNN \cite{pons2017timbre} & 89.30  & - & 191k  \\
Timbre CNN {(reproduced)}\footnotemark[14] & 89.07 & 34.92 & 220k \\	
\multicolumn{4}{l}{\textit{}} \\
\multicolumn{4}{l}{\textit{How much capacity is required for the front-end?}} \\\hline
\# filters $\times$1/8	& 90.08 &37.18 & 4.4M \\
\# filters $\times$1/4	& 90.12& 37.69 & 4.6M \\
\# filters $\times$1/2\hspace{6mm}(selected)	& \textbf{90.40}& \textbf{38.11} & 5M \\
\# filters $\times$1& 90.31&37.79 & 5.9 \\
\# filters $\times$2& 90.07	& 37.29& 7.6M  \\	
\multicolumn{4}{l}{\textit{}} \\
\multicolumn{4}{l}{\textit{How much capacity is required for the back-end?}} \\
\multicolumn{4}{l}{\# filters in every CNN layer - \# units in dense layer} \\
\hline
\hspace{1.2mm} 64 CNN filters - 500 units  &90.03&36.98  & 277k\\
\hspace{23.5mm}- 200 units  & \underline{90.28}&\underline{37.55}& 222k\\
128 CNN filters - 500 units  & 90.16&37.61&617k
 \\
\hspace{22.5mm} - 200 units  & \underline{90.28}&\underline{37.69}&524k \\
256 CNN filters - 500 units  &90.18&37.98&1.6M \\
\hspace{22.5mm} - 200 units  &90.06&37.16&1.4M \\
512 CNN filters - 500 units &\underline{90.40}&\underline{38.11}& 5M \\
\hspace{22.5mm} - 200 units  &89.98&37.05 & 4.7M\\

	\end{tabular} 
			\vspace{-2mm}
			\caption{MTT results: spectrogram models.}
			\label{tab:spec}
			\vspace{-0mm}
\end{table}

\vspace{-2mm}
\subsection{Million Song Dataset (MSD)}
\vspace{-1mm}
\hspace{3mm} \textbf{Experimental setup.} State-of-the-art models are set as baselines, and we use the same (classification) performance metrics as for the 1.2M-songs dataset: ROC-AUC and PR-AUC -- note that the MSD labels are binary. 
These experiments aim to validate the studied models with the biggest public dataset available. 
Models are set following the \textit{basic} configuration, and results are shown in Table \ref{tab:msd}.

\textbf{Quantitative results.} 
The spectrogram model outperforms the waveform model for this public dataset -- having $\approx$~200k training songs. Furthermore, the spectrogram model performs equivalently to `Multi-level \& multi-scale'~\cite{lee2017multi}, which is the best performing method in the literature -- denoting that musical knowledge can be of utility to design models for the MSD. Additionally, the waveform model performs worse than other waveform-based models that also employ sample-level front-ends.
Such performance decrease could be caused because \textit{(i)} SampleCNN methods \cite{lee2017sample,lee2018samplecnn} average ten\footnote{Since MSD audios are of 30s, ten tag estimates per song can be obtained via running the model with consecutive patches of~3s.}estimates for the same song to compensate for possible faults in song-level predictions, while our method only averages two -- predicting consecutive patches of 15s; or \textit{(ii)}~because the major difference between SampleCNN and the waveform model is that the latter employs a global pooling strategy that could remove potentially useful information for the model.
Besides, the best performing waveform-based model (`SampleCNN~multi-level \& multi-scale'~\cite{lee2018samplecnn}) also achieves lower scores than the best performing spectrogram-based ones. Considering the outstanding results we report when the  waveform model is trained with 1M songs, one could argue that the lack of larger public datasets is limiting the outcomes of deep learning research for music auto-tagging -- particularly when dealing  with end-to-end learning stacks processing raw data such as waveforms.


\begin{table}[b]
	\vspace{-4mm}
	\centering
	\begin{tabular}{l||c c||c}
		\textbf{MSD} & \textit{ROC} & \textit{PR} & \# \\ 
		\textit{Models}  & \textit{AUC} & \textit{AUC} & \textit{param} \\ \hline \hline
		{Waveform} \hspace{0.5mm} \textit{(ours)} & 87.41&28.53  & 5.3M \\\hline
		SampleCNN \cite{lee2017sample}	& {88.12} & - & 2.4M\\\hline
		SampleCNN multi-level& \multirow{2}{*}{\textbf{88.42}} & \multirow{2}{*}{-} & \multirow{2}{*}{-} \\
		\& multi-scale\cite{lee2018samplecnn} &  &  &  \\	
		\hline \hline
		{Spectrogram} \hspace{0.5mm} \textit{(ours)} & \textbf{88.75} & \textbf{31.24} & 5.9M \\ \hline
		VGG + RNN \cite{choi2017convolutional}	& 86.2 & - & 3M\\ \hline
		Multi-level \& & \multirow{2}{*}{\textbf{88.78}} & \multirow{2}{*}{-} & \multirow{2}{*}{-} \\
		multi-scale\cite{lee2017multi} &  &  &  \\											 		 
	\end{tabular} 
	\vspace{-0mm}
	\centering
	\caption{MSD results. \textbf{\textit{Top}} -- waveform-based models.\newline \textbf{\textit{Bottom}} -- spectrogram-based models. }
	\label{tab:msd}
\end{table}
\footnotetext[13]{Result computed with a different MTT version, see section 5.2.}
\footnotetext[14]{Reproduced using 96 mel bands instead of 128 as in \cite{pons2017timbre}.}
\vfill\eject
\phantom{}
\vspace{-14.5mm}
\section{Conclusions}

This study presents the first work describing how different deep music auto-tagging architectures perform depending on the amount of available training data. We also present two architectures that yield results on par with the state-of-the-art.
These architectures are based on two conceptually different design principles: one is based on a waveform front-end, and no domain knowledge inspired its design; 
and the other, with a spectrogram front-end, makes use of (musical) domain knowledge to justify its architectural choices.  
While our results suggest that models relying on domain knowledge play a relevant role in scenarios where no sizable datasets are available, we have shown that, given enough data, assumption-free models processing waveforms outperform those that rely on musical domain knowledge.





\section{Acknowledgments}
 
This work was partially supported by the Maria de Maeztu Units of Excellence Programme (MDM-2015-0502) -- and we are grateful for the GPUs donated by NVidia.

\bibliography{ISMIRtemplate}

%
%
%
%

\end{document}